\documentclass{pasj00}
\usepackage[dvips]{graphicx}

\begin{document}
\SetRunningHead{Nakagawa et al.}{Astrometry of T~Lep with VERA}
\Received{2000/12/31}%{yyyy/mm/dd}
\Accepted{2001/01/01}%{yyyy/mm/dd}

\title{VLBI Astrometry of AGB Variables with VERA\\
      -- A Mira Type Variable T~Lepus --}

\author{
Akiharu \textsc{Nakagawa}$^{1}$, 
Toshihiro \textsc{Omodaka}$^{1}$, 
Toshihiro \textsc{Handa}$^{1}$, 
Mareki \textsc{Honma}$^{2}$, \\
Noriyuki \textsc{Kawaguchi}$^{3}$,
Hideyuki \textsc{Kobayashi}$^{2}$, 
Tomoaki \textsc{Oyama}$^{2}$, 
Katsuhisa \textsc{Sato}$^{3}$, \\
Katsunori~M. \textsc{Shibata}$^{1}$,  
Makoto \textsc{Shizugami}$^{3}$, 
Yoshiaki \textsc{Tamura}$^{3}$, and 
Yuji \textsc{Ueno}$^{3}$
}
\affil{$^{1}$Graduate School of Science and Engineering, Kagoshima University, \\
1-21-35 Korimoto, Kagoshima-shi, Kagoshima 890-0065}
\affil{$^{2}$Mizusawa VLBI Observatory, National Astronomical Observatory of Japan, 
2-21-1 Osawa, Mitaka, Tokyo 181-8588}
\affil{$^{3}$Mizusawa VLBI Observatory, National Astronomical Observatory of Japan, \\
2-12 Hoshi-ga-oka, Mizusawa-ku, Oshu-shi, Iwate 023-0861}
\email{nakagawa@sci.kagoshima-u.ac.jp}

%% `\KeyWords{}' always has to be placed before `\maketitle'.
\maketitle

\begin{abstract}
We conducted phase referencing VLBI observations of the Mira variable T~Lepus (T~Lep) using VERA, from 2003 to 2006. 
The distance to the source was determined from its annual parallax which was measured to be 3.06$\pm$0.04 mas, corresponding to a distance of 327$\pm$4\,pc. 
Our observations revealed the distribution and internal kinematics of H$_2$O masers in T~Lep, and 
we derived a source systemic motions of 
14.60$\pm$0.50 mas\,yr$^{-1}$ and  $-$35.43$\pm$0.79 mas\,yr$^{-1}$ 
in right ascension and declination, respectively. 
We also determined a LSR velocity of $V_\mathrm{LSR}^{\ast} = -$27.63 km\,s$^{-1}$.  
Comparison of our result with an image recently obtained from the VLTI infrared interferometer reveals a linear scale picture of the circumstellar structure of T~Lep. 
Analysis of the source systemic motion in the Galacto-centric coordinate frame indicates a large peculiar motion, which is consistent with the general characteristics of AGB stars. 
This source makes a contribution to the calibration of the period-luminosity relation of Galactic Mira variables. 
From the compilation of data for nearby Mira variables found in the literature,  whose distances were derived from astrometric VLBI observations, we have calibrated the Galactic Mira period-luminosity relation to a high degree of accuracy.  
\end{abstract} 

\section{Introduction} 
Mira variables are pulsating stars of masses 1--8\,M$_\odot$ which undergo rapid mass loss before ejecting their outer layers as planetary nebula shells. They play an important role in the studies of evolved stars and are active sources of chemical enrichment in the universe. Mira variables show a relation between their pulsation period and luminosity, the so called ``Period--luminosity relation (PLR)''. 
One of the goals of the VERA project (VLBI Exploration of Radio Astrometry) is to determine the PLR of Galactic Mira variables. 
VERA is a VLBI  array dedicated to the study of Galactic dynamics \citep{kob03}. 
Calibration of the PLR for Galactic Mira variables is important because the PLR is used to determine the distances of Mira variables based on their pulsation periods and apparent magnitudes.  
On this basis, the distribution and kinematics of Mira variables is used to study the physics of the evolved star component of the Galaxy.
As T~Lepus (T~Lep) is a Mira variable with a pulsation period of 368\,days (GCVS)\footnote[1]{General Catalog of Variable Stars\\http://heasarc.gsfc.nasa.gov/W3Browse/all/gcvs.html}
 and accompanied with bright H$_2$O masers, it is a fitting target for our study. 
T~Lep has a spectral type of M6e--M9e, and its V-band magnitude varies from 7.3 to 14.3 mag \citep{sam04}. 
From the 2MASS database, the $K$-band magnitude is reported to be $-$0.266$\pm$0.354 mag \citep{cut03}, and the mass-loss rate was estimated by \citet{lou93} to be 7.3$\times$10$^{-7}$ M$_\odot$yr$^{-1}$. 
In the first Hipparcos catalog \citep{per97}, the parallax of T~Lep is reported to be negative ($-$1.77$\pm$2.73 mas), and its distance can not be derived. 
Recently, new reduction of the Hipprarcos data \citep{van07} provides a parallax of 1.37$\pm$1.44 mas. 
Although the nominal parallax becomes positive, the error is still large and thus the corresponding distance of $\sim$730 pc is not reliable, with roughly $\pm$ 100\% error. 
Therefore, determination of a more accurate distance is desirable. 

Observations of the photosphere and molecular shells of AGB stars have seen significant improvements recently in the form of infrared interferometry, and T Lep is one of the few sources observed in detail in this way.
With an accurate distance, we can convert the angular size obtained from the infrared interferometer to the linear size of the star. 
The lack of a reliable distance estimate prevents us from understanding the physical properties of stars. 
By measuring the trigonometric parallax with VERA, we are able to evaluate the parameters to a higher accuracy than those previously obtained. 
In this paper, we report distance and proper motion of T~Lep, measured with the astrometric VLBI observations. 
With regards to the calibration of the PLR for Galactic Mira variables, the number of accurate astrometric results is still small. 
Calibration of the Galactic PLR is one of the principle aims of our study and our parallax measurement of T~Lep contributes an additional good source to the PLR sample used to investigate and improve the reliability of the PLR. 

Details of our observations and data reduction are described in section~\ref{sec_obs_red}. 
In section~\ref{results}, we present the motions of maser spots revealed by multi epoch VLBI observations with VERA. 
The internal kinematics of the masers associated with T~Lep are also presented. 
Trigonometric parallax is converted to distance in this section. 
In section~\ref{discussion}, we estimate position of photosphere of T Lep using the observed distribution and internal kinematics of masers. 
Furthermore, we calculate the spatial motion of T Lep in the Galacto-centric coordinate frame, and also discuss the calibration of the Galactic PLR in section~\ref{discussion}.
We summarize our study in section~5. 

\section{Observations and data reduction}
\label{sec_obs_red}
\subsection{Observations}
\label{subsec_obs} 
Between February 2004 and March 2008, a total of 20 astrometric VLBI observations toward T~Lep were performed with VERA. 
We observed the H$_2$O maser emission at 22.235 GHz.
Using the dual-beam system of VERA \citep{kaw00}, an extragalactic continuum source J0513$-$2159
was simultaneously observed at 22 GHz as a position reference for the phase-referenced images. 
The separation angle of T~Lep and J0513$-$2159 was $2.08^{\circ}$ with a pair position angle of $92^{\circ}$. 
The typical duration of each observation was 6--8 hours, in order to track the sources from horizon to horizon, and the observation interval was approximately one month. 

In our observations we received left handed circularly polarized emission.
Two data recording rates of 128 Mbps and 1024 Mbps in 2-bit digitization were adopted, depending on the observing epoch. 
In 2004 and 2005, 10 observations were recorded in 128 Mbps mode, and the remaining 10 observations were carried out in 1024 Mbps mode. 
The recording rates of 128 Mbps and 1024 Mbps yield total bandwidths of 32 MHz and 256 MHz, respectively. 
In the observations utilizing a 128 Mbps recording rate, the 32 MHz band width data was divided into 2 IF channels of 16 MHz bandwidth, with one IF channel being used to receive maser emission form T~Lep and the other used to receive continuum emission from J0513$-$2159.
In the observations utilizing a 1024 Mbps recording rate, the total bandwidth of 256 MHz was divided into 16 IF channels of 16 MHz bandwidth. 
Then, one IF was used for T~Lep and the other 15 IFs were used for J0513$-$2159. 

In 19 observations, 
we adopted a frequency spacing of 15.625 kHz corresponding to a velocity spacing of 0.21 km\,s$^{-1}$. 
In the observation on 2006 Mar 20, a frequency spacing of 32 kHz was adopted, yielding a velocity spacing of 0.42 km\,s$^{-1}$. 
In the case of the reference source data from J0513-2159, each IF channel was divided into 64 spectral channels.

The \textit{a priori} coordinates of the two sources in J2000 equinox were
($\alpha$, $\delta$) $=$ 
(05$^\mathrm{h}$ 04$^\mathrm{m}$ 50$^\mathrm{s}$.843,      $-$21$^{\circ}$ 54$\arcmin$ 16$\arcsec$.505) for T~Lep and 
(05$^\mathrm{h}$ 13$^\mathrm{m}$ 49$^\mathrm{s}$.114324, $-$21$^{\circ}$ 59$\arcmin$ 16$\arcsec$.09203) for J0513$-$2159, respectively. 
These coordinates are adopted as phase tracking centers in the correlation process using the Mitaka FX correlator~\citep{shi98} at the National Astronomical Observatory of Japan (NAOJ). 
The typical synthesized beam size (FWHM) was 1.70 mas $\times$ 0.66 mas with a position angle of $160^{\circ}$. 
Table~\ref{table_obs} summarizes the observations carried out with VERA. 
The times of the observations are given as Date (year, month, day), Modified Julian Date (MJD), and Day of Year (DOY). 
The recording rates are also presented.

In parallel with the VLBI observations, we have monitored H$_2$O maser emission in T~Lep with single-dish observations at the VERA Iriki station. 
The single-dish monitoring started in 2003 September with a typical interval of one month \citep{shi08}. 
The 1-$\sigma$ noise level of the single-dish observations is 0.05 K, corresponding to $\sim$1 Jy. 

\begin{table}[!tb]
\caption{VLBI Observations.}
\label{table_obs}
\begin{center}
\begin{tabular}{crcrrr} 
\hline
Epoch & \multicolumn{1}{c}{Date} & MJD & DOY & Rate. \\ \cline{2-2}
ID\footnotemark[$\dag$] &     &   &   & [Mbps] \\ \hline \hline 
1$^{\,\,\,\,}$& 2004\,\,\,Feb  02  & 53037 & 33 & 1024 \\
2$^{\,\,\,\,}$&       Mar    01  & 53065 & 61 &  128 \\
3$^{\,\,\,\,}$&       Mar    25  & 53089 & 85 &  128 \\
4$^{\,\,\,\,}$&       Apr    26  & 53121& 117 &  128 \\
5$^{\,\,\,\,}$&       May   19  & 53144 & 140 & 1024 \\
6$^{\,\,\,\,}$&       Nov    24  & 53333 & 329 &  128 \\
7$^{\ast}$&      Dec    26  & 53565 & 360 &  128 \\ \hline 
8$^{\ast}$  & 2005\,\,\,Jan   19  & 53389 & 19 &  128 \\
9$^{\ast}$  &      Feb   11  & 53412 & 42 &  128 \\
10$^{\ast}$  &       Mar   10  & 53439 & 69 &  128 \\
11$^{\,\,\,\,}$&       Apr   11  & 53471 & 101 &  128 \\
12$^{\,\,\,\,}$&       May   12  & 53502 & 132 &  128 \\
13$^{\ast}$  &       Sep   20  & 53633 & 263 & 1024 \\
14$^{\ast}$  &       Nov   21  & 53695 & 325 & 1024 \\
15$^{\ast}$  &       Dec   22  & 53726 & 356 & 1024 \\ \hline 
16$^{\,\,\,\,}$& 2006\,\,\,Jan   31 & 53766 & 31 & 1024 \\
17$^{\ast}$  &       Mar  20  & 53814 & 79 & 1024 \\
18$^{\ast}$  &       Apr   20  & 53845 & 110 & 1024 \\
19$^{\,\,\,\,}$&       May   21  & 53876 & 141 & 1024 \\
20$^{\,\,\,\,}$&       Aug  10  & 53957 & 222 & 1024 \\
\hline   
\multicolumn{5}{@{}l@{}}{\hbox to 0pt{\parbox{77mm}{\footnotesize
\smallskip
\par\noindent
\footnotemark[$\dag$] 
Epoch IDs marked with ``$\ast$'' indicate epochs that were used in the parallax estimation.
 }\hss}}
\end{tabular} 
\end{center}  
\end{table}   

\subsection{Data Reduction} 
\label{datareduction}
\subsubsection{Phase-referencing imaging} 
\label{subsub_phas}
For the determination of an annual parallax and proper motion of maser spots, we must measure their positions with respect to the reference source J0513$-$2159 via a data reduction process incorporating phase-referencing. 
In data reduction and imaging processes, we used the Astronomical Image Processing System (AIPS) package. 
The amplitude calibration was performed by using system temperatures and antenna gains logged during the observations. 
To solve residual phase fluctuation in the raw data of reference source J0513$-$2159, we used the task {\sc fring} with an integration time of 2$-$4 minutes. 
Solutions of phases, group delays, and delay rates obtained in every 30 seconds were transferred to the data of T~Lep to calibrate its residual phase. 
Calibrated visibility data of T~Lep were finally Fourier$-$transformed to create phase-referenced images with dimensions 1024 $\times$ 1024 pixels, with a pixel size of 0.05 mas. 
The brightness distribution on the final images was fitted to two dimensional Gaussian models to obtain positions of the maser spots. 
Then, the positions defined as the peak of the Gaussian models were used to estimate the annual parallax and linear proper motions. 
The 1-$\sigma$ noise level of phase-referenced images ranges from 0.2 Jy\,beam$^{-1}$  to 2.0 Jy\,beam$^{-1}$ and the lowest signal-to-noise ratio (S/N) of any phase-referenced image was 6.7, on 2005 Feb 11. 

\subsubsection{Single-beam VLBI imaging} 
\label{subsub_single}
In addition to the phase-referencing imaging, we also performed single-beam VLBI imaging of maser spots in order to obtain their distribution and the internal maser kinematics in T~Lep. 
We used the same software package. 
In this reduction procedure we only solved for group delay, this was done using the bright continuum source J0530$+$1331.
Then, residual phases and rates were solved using a bright maser spot with LSR velocity ($V_{\mathrm{LSR}}$) of $-$29.73 km\,s$^{-1}$. 
In the single-beam VLBI images, the positions of other maser spots were determined with respect to this reference maser spot.
In deriving the internal motions of maser spots, a linear least-squares analysis was applied to maser spots that were detected at the same velocity channel during at least two continuous observations. 
The 1-$\sigma$ noise level of the single-beam VLBI images ranges from 130 mJy\,beam$^{-1}$ to 590 mJy\,beam$^{-1}$, and a S/N of 7 was adopted as the detection criterion. 

\section{Results}
\label{results}
\subsection{Single-dish observation results}
\label{maser_spectrum}
\begin{figure}[tb]
\begin{center}
 \includegraphics[width=87mm, angle=0]{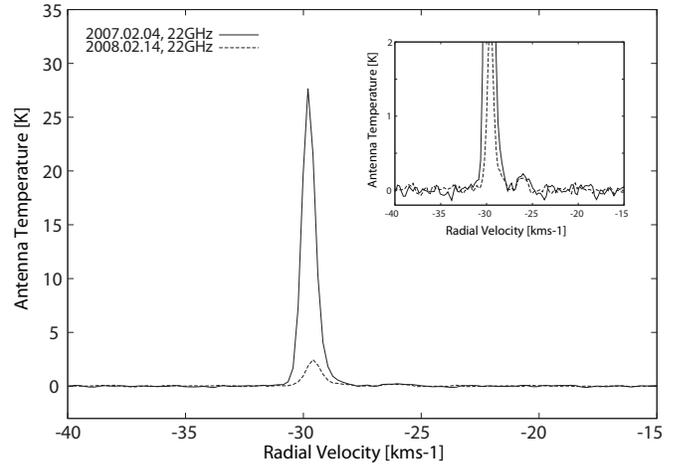}
\end{center}
 \caption{
Total-power spectra of H$_2$O masers in T~Lep observed at the Iriki station in 2007 Feb 4 (solid line) and 2008 Feb 14 (dashed line). 
The profile of a weak emission at $V_{\mathrm{LSR}}$ of $-26$ km\,s$^{-1}$ can be seen in the magnified inset. The velocity component at $V_{\mathrm{LSR}}$ of $-30$km\,s$^{-1}$ was persistently dominant. 
}
 \label{fig-spectr}
\end{figure}

Total-power spectra of H$_2$O masers in T~Lep is presented in figure~\ref{fig-spectr}. 
The spectra were obtained with single-dish observations at VERA  Iriki station on 2007 Feb 4 (solid line) and 2008 Feb 14 (dashed line).
During our monitoring period, a single velocity component at $V_{\mathrm{LSR}}$ of $-30$ km\,s$^{-1}$ was dominant. 
We magnified an inset in figure~\ref{fig-spectr} to show weak emission at $V_{\mathrm{LSR}}$ of $-26$ km\,s$^{-1}$.

\begin{figure*}[htpb]
\begin{center} 
 \includegraphics[width=130mm, angle=0]{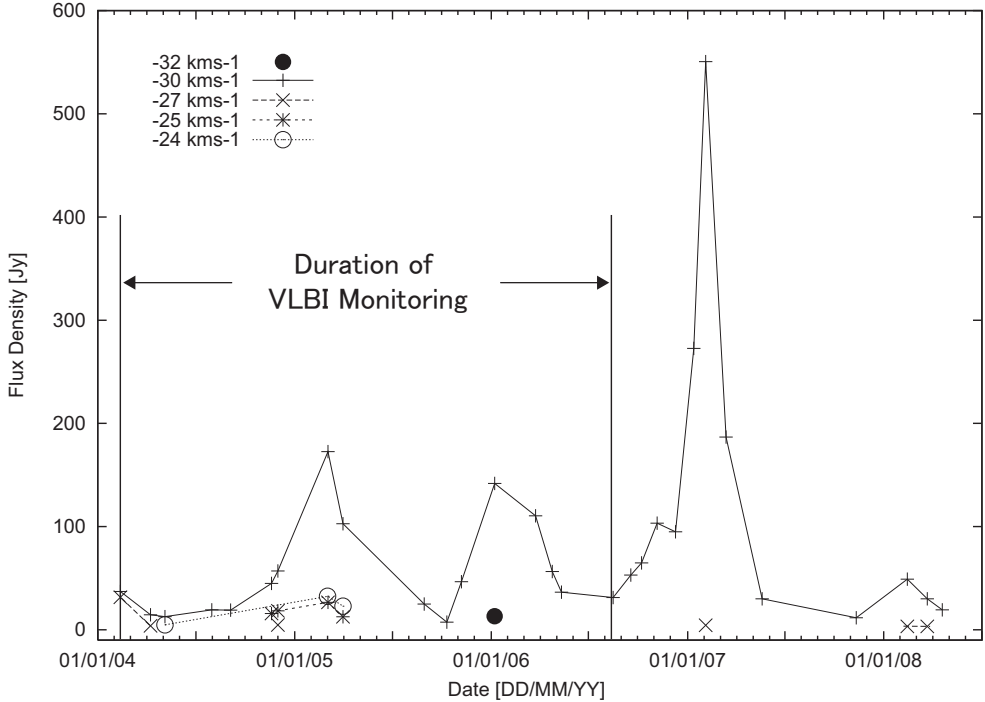}
\end{center}  
 \caption{
Time variation of H$_2$O masers obtained with single-dish monitoring at the VERA Iriki station. 
Major velocity components are presented with different symbols 
of ``{\Large$\bullet$}'', ``$+$'', ``$\times$'', ``{\large $\ast$}'', and ``{\Large$\circ$}'', corresponding to $V_{\mathrm{LSR}}$ of $-32$, $-30$, $-27$, $-25$, and $-24$ km\,s$^{-1}$, respectively. 
Time duration of our VLBI observations is also presented. 
}
\label{fig-lc}
\end{figure*}

Time variation of the maser over four years is presented in figure~\ref{fig-lc}, with the intensity variation of five velocity components denoted by different symbols. 
The component with a $V_{\mathrm{LSR}}$ of 
$-30$ km\,s$^{-1}$ was bright and continuously detected during our single-dish monitoring observations. 
A pulsation period of 368 days (GCVS), derived from optical observations, seems to be consistent with the radio intensity variation. 
The period of our VLBI observations is also shown in figure~\ref{fig-lc}. 
Intensities of the other velocity components at $V_{\mathrm{LSR}}$ of $-24$ km\,s$^{-1}$, $-25$ km\,s$^{-1}$, $-27$ km\,s$^{-1}$, and $-32$ km\,s$^{-1}$ were much weaker, and could be detected only a limited number of epochs in our single-dish observations. 

\subsection{Annual parallax and distance}
\label{parallax}
Using phase-referenced images we estimated the parallax of T Lep.
During our observation periods, the maser spot at $V_{\mathrm{LSR}}=$ $-29.73$ km\,s$^{-1}$ was persistently brighter than all other maser spots over 1.5 years, and detected in phase-referenced images in 9 epochs (epoch IDs of 7, 8, 9, 10, 13, 14, 15, 17, and 18). 
Intensity of this maser spot in phase-referenced images showed time variation between 1.5 Jy\,beam$^{-1}$ on 2005 Feb 11 and 19.0 Jy\,beam$^{-1}$ on 2005 Sep 20, with corresponding S/Ns of 6.7 and 9.7, respectively. 
The averaged S/N of phase-referenced images was 9. 
Using the maser spot at $V_{\mathrm{LSR}}=$ $-29.73$ km\,s$^{-1}$, we determined the annual parallax and linear proper motion. 
The resultant annual parallax was estimated to be 3.06$\pm$0.04 mas, corresponding to a distance of 327$\pm$4 pc. 
In this fitting process, position uncertainties in R.A.($\sigma_{\mathrm{X}}$) and Dec.($\sigma_{\mathrm{Y}}$) of 0.08 mas and 0.18 mas were applied. 
These position uncertainties were obtained as a quadratic sum of possible error factors. 
More details are presented in section~\ref{error}. 

The linear proper motions of the maser spot at $V_\mathrm{LSR}=-29.73$ km\,s$^{-1}$ along the R.A. and Dec. axes ($\mu_{\mathrm{X}}$, $\mu_{\mathrm{Y}}$) were determined to be ($13.59\pm 0.05$, $-34.55\pm 0.12$) mas\,yr$^{-1}$. 
This motion is a combination of systemic motion of T~Lep and the internal motion component of the spot with respect to the system. 

\begin{figure}[tb]
\begin{center}
\includegraphics[width=87mm, angle=0]{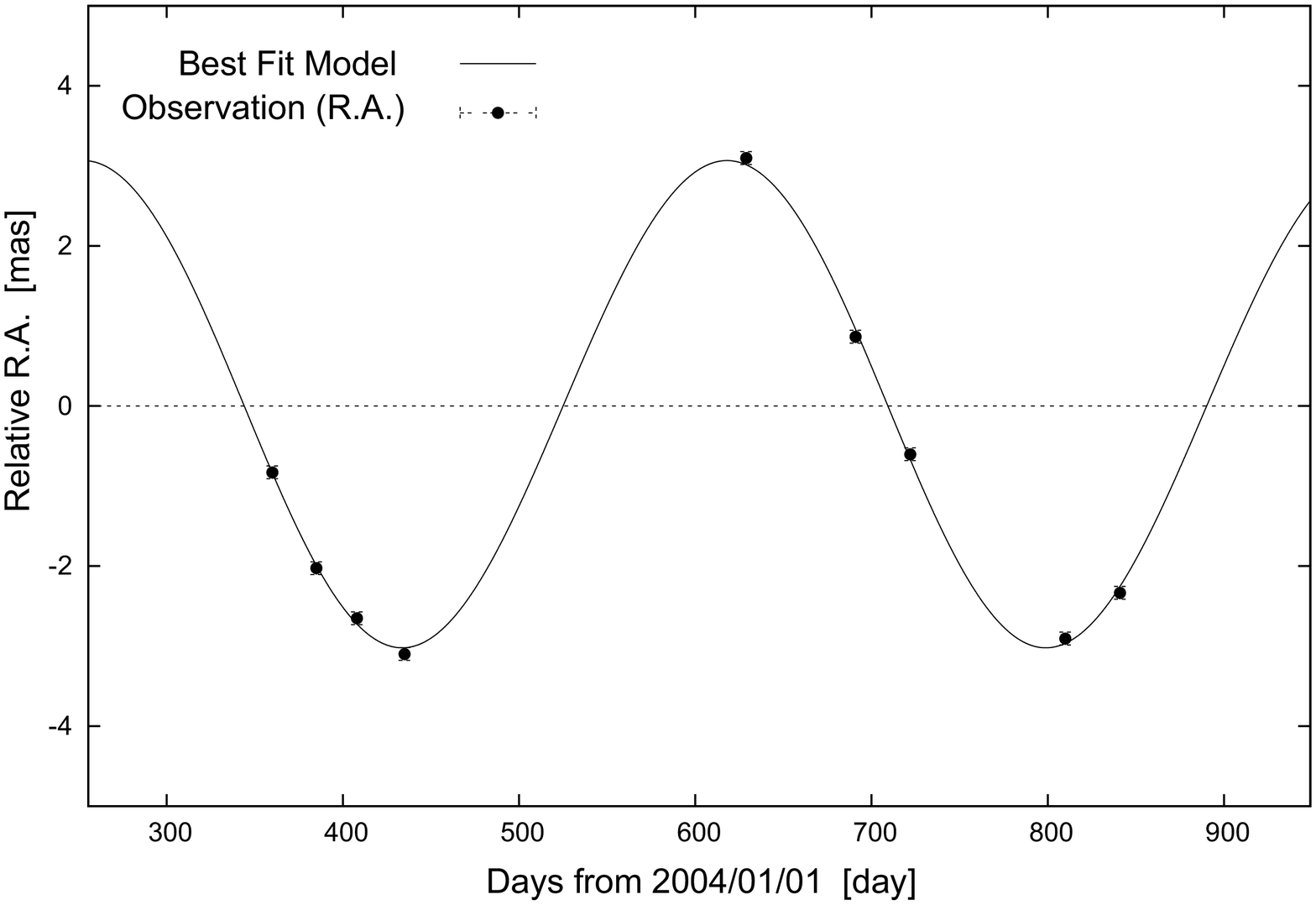} 
\includegraphics[width=87mm, angle=0]{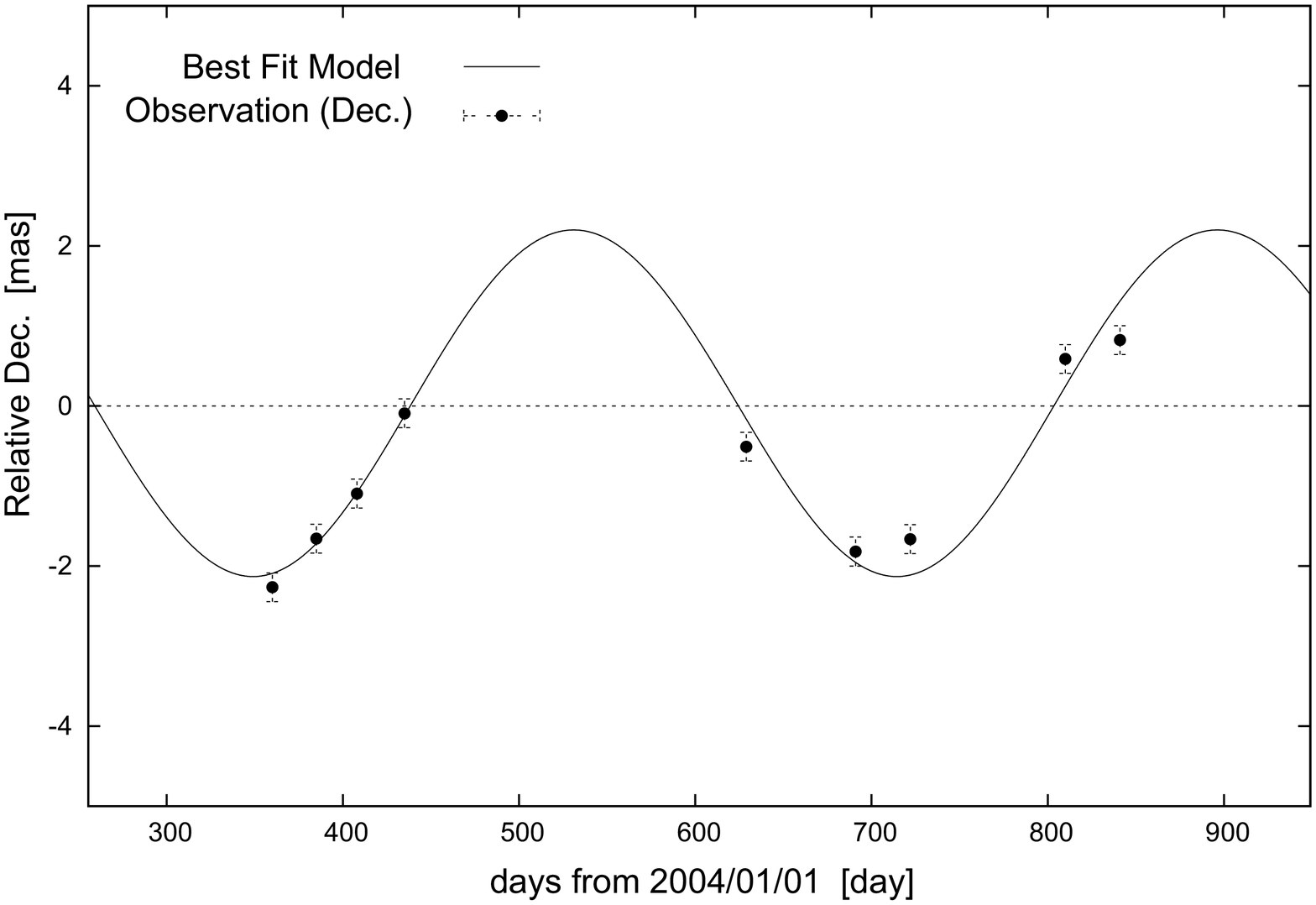} 
\end{center}
\caption{
Oscillation of the maser spot at $V_{\mathrm{LSR}} = -29.73$ km\,s$^{-1}$ in T~Lep. 
Filled circles represent the maser position obtained by phase-referencing analysis. 
Solid lines indicates the best fit model of the parallactic oscillation. 
}
\label{fig_parallax}
\end{figure}

In figure~\ref{fig_parallax}, we present parallactic oscillations of the maser along R.A. (top panel) and Dec. (bottom panel) axes. 
Horizontal axis indicates the days from 2004 Jan 1. 
Each vertical axis shows positional offset after subtracting the linear proper motion. 
Error bars indicate position uncertainties along each axis. 
Sizes of the error bars in R.A. is almost the same as the size of the symbol. 

The image of J0513$-$2159 did not show any change of the structure and it was always observed as a point source, confirming that J0513$-$2159 wad suitable as a position reference throughout our observations. 

\subsection{Astrometric Error}\label{error}
In phase referencing observations, there are several factors that cause positional error in the image of the maser.  
In this section, we estimate the positional error due to these factors. 
At first, we consider the error which depends on the S/N of the phase-referenced image (thermal noise). 
From the image S/N and synthesized beam size $\theta_{\rm{b}}$, the positional error is estimated as $\theta_{\rm{b}}/(\rm{S/N}$) which, using our average S/N of 9, corresponds to errors of 65\,$\mu$\,as and 178\,$\mu$\,as in R.A. and Dec., respectively.

Next, we estimate the error contribution from residuals of the wet zenith excess path. 
In our previous study \citep{nak08}, the typical error of this kind was estimated to be $\sim$30 mm. 
Based on the same procedure used in section~4 in \citet{hon07}, we estimated a positional error of 45 $\mu$as. 
Since the source pair position angle is $92^{\circ}$, the difference of elevation angles is $\sim1^{\circ}$ at the time of the setting and rising of the source pair. 
This causes a path-length error of 0.5 mm ($=$\,30 mm$\times$1$^{\circ}$/57$^{\circ}$.3/rad, where 1$^{\circ}$ is the elevation angle difference of the pair) between the two sources. 
This path length error corresponds to 45\,$\mu$\,as ($=$ 0.5 mm/2.3$\times 10^9$ mm where 2.3$\times 10^9$ mm is the maximum baseline length of VERA). 
Although the error in the antenna position of each station gives another astrometric error, this contribution is one order of magnitude smaller than that due to the wet zenith excess path. 

Taking a quadratic sum of these factors, we finally obtained the position uncertainties in R.A. ($\sigma_{\mathrm{X}}$) and Dec. ($\sigma_{\mathrm{Y}}$) to be 0.08 mas and 0.18 mas, respectively. 
We use these errors in the estimation of the annual parallax. 

\subsection{Maser distribution and internal motions}
\label{distribution}

\begin{figure*}[tb]
\begin{center}
 \includegraphics[width=120mm, angle=0]{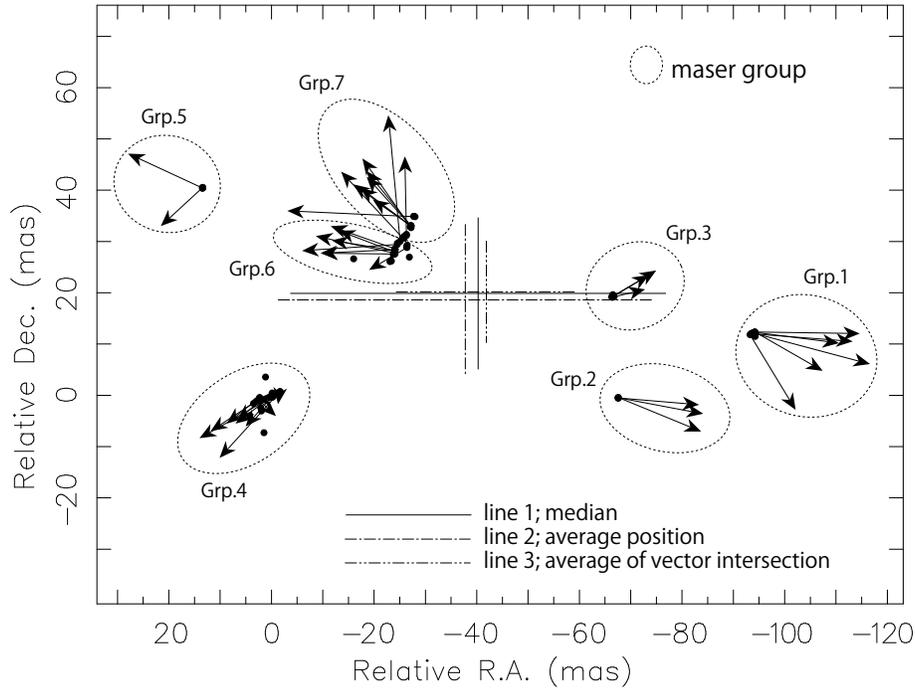}
\end{center}
\caption{
Illustration of the maser groups and estimated position of the central star. 
H$_2$O masers were divided into 7 groups and presented by circles with dotted line. 
Estimated star positions are presented by crosses using different line styles. 
See section~\ref{star_position} for detail. 
Arrows are same as those in figure~\ref{map}
}
 \label{map_grp}
\end{figure*}

\begin{figure*}[tb]
\begin{center}
 \includegraphics[width=150mm, angle=0]{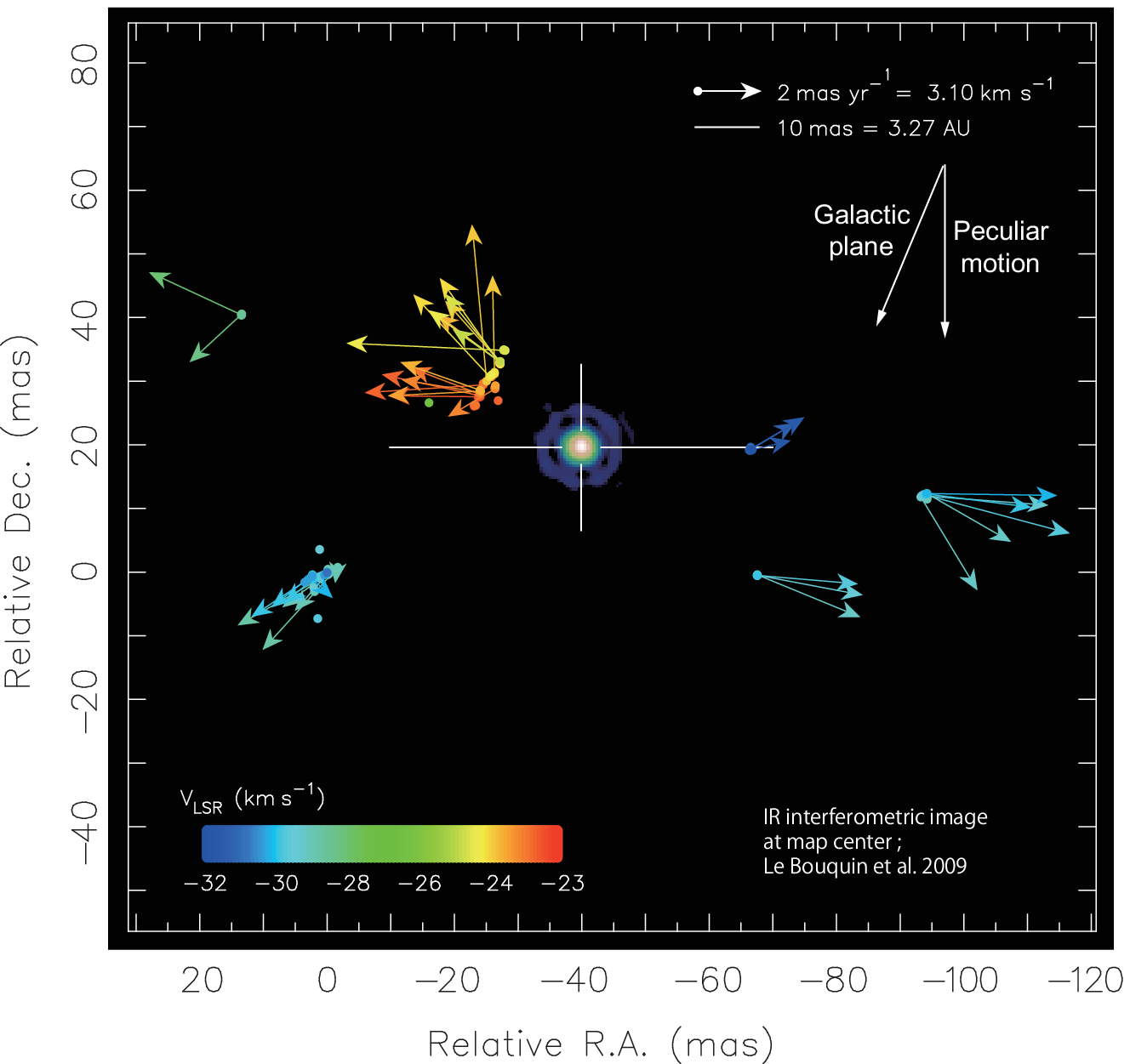}
\end{center}
\caption{
H$_2$O maser distribution in T~Lep. 
Superposition of VLTI infrared interferometric image and H$_2$O maser distribution obtained with VERA. 
Color image at the center of this figure is an image obtained with VLTI. 
H$_2$O masers observed with VERA are distributed at outer area of this figure. 
Colors of filled circles indicate their $V_{\mathrm{LSR}}$. 
Color of the central star is unrelated to the color index of the maser.
}
 \label{map}
\end{figure*}

To investigate the distribution and internal kinematic of masers in T~Lep, we used images obtained using the single-beam VLBI imaging process as described in section~\ref{subsub_single}. 
We detected 84 maser spots toward T Lep, these are listed in table~\ref{maser_table}, in order of increasing $V_{\mathrm{LSR}}$. 

Regarding the time variation of maser positions, we consider the internal kinematics in the T~Lep system as follows. 
At first, we obtained the linear motion of $i$-th maser spot ($v_{\mathrm{x}}^i$, $v_{\mathrm{y}}^i$), in units of mas\,yr$^{-1}$, relative to the reference maser spot (ID 57). 
The subscripts of $\mathrm{x}$ and $\mathrm{y}$ denote directions along the R.A. and Dec., respectively. 
The degree of error in linear motions were also obtained from this fitting. 
Based on this data, we defined a ``maser group'' as a collection of spots for which the $V_\mathrm{LSR}$ differences were less than $\sim$1 km\,s$^{-1}$ and the position differences on sky plane were within $\sim$5 mas.
Under these criteria we found seven groups of maser spots and an isolated spot (ID 35). 
The spot ID\,35 exhibits large differences in both position and  $V_{\mathrm{LSR}}$. 
Each group is composed of a few to several maser spots of proximal location and similar $V_{\mathrm{LSR}}$, these are represented by dotted circles in figure~\ref{map_grp}. 
The group IDs are shown as ``Grp.1'' to ``Grp.7'' in figure~\ref{map_grp} and also in column (11) of table~\ref{maser_table}.
Then we obtained proper motion of the $j$-th group ($v_{\mathrm{x}}^j$, $v_{\mathrm{y}}^j$) by averaging the proper motions as, 
\begin{eqnarray}
 v_{\mathrm{x}}^j ~= ~ \frac{1}{n} \sum_{i=1}^{n} v_{\mathrm{x}}^i,\,\,\,\,\,\,\,\,\,\,
 v_{\mathrm{y}}^j ~= ~ \frac{1}{n} \sum_{i=1}^{n} v_{\mathrm{y}}^i
\label{x-offset}
\end{eqnarray}
where, $n$ is the number of maser spots included in the groups $j$. 

Next, we estimate the mean motions 
($V_{\mathrm{x}}^{\mathrm{mean}}$, $V_{\mathrm{y}}^{\mathrm{mean}}$) 
from the following, 
\begin{eqnarray}
 V_{\mathrm{x}}^{\mathrm{mean}} ~= ~ \frac{1}{7} \sum_{j=1}^{7} v_{\mathrm{x}}^j,\,\,\,\,\,\,\,\,\,\,
 V_{\mathrm{y}}^{\mathrm{mean}} ~= ~ \frac{1}{7} \sum_{j=1}^{7} v_{\mathrm{y}}^j,  
\label{x-offset}
\end{eqnarray}
where, 7 is the number of maser groups. 
Using the observed data, the mean motion ($V_{\mathrm{x}}^{\mathrm{mean}}$, $V_{\mathrm{y}}^{\mathrm{mean}}$) $=$ ($-$1.01, 0.88) mas\,yr$^{-1}$ was obtained. 
After subtracting this mean motion from the previously obtained maser motions  
($v_{\mathrm{x}}^i$, $v_{\mathrm{y}}^i$), we finally obtain the internal motions 
($V_{\mathrm{x}}^i$, $V_{\mathrm{y}}^i$) of each maser. 

\begin{eqnarray}
 V_{\mathrm{x}}^i = v_{\mathrm{x}}^i - V_{\mathrm{x}}^{\mathrm{mean}} ,\,\,\,\,\,\,\,\,\,\,
 V_{\mathrm{y}}^i = v_{\mathrm{y}}^i - V_{\mathrm{y}}^{\mathrm{mean}}  
\label{xy_int}
\end{eqnarray}

These relative internal motions ($V_{\mathrm{x}}^i$, $V_{\mathrm{y}}^i$) are represented by arrows in figure~\ref{map}. 
For the error values of $V_{\mathrm{x}}^i$ and $V_{\mathrm{y}}^i$, we adopted the same values obtained in the least-squares fitting of $v_{\mathrm{x}}^i$ and $v_{\mathrm{y}}^i$. 
For 84 maser spots, the positions ($X^i$, $Y^i$), internal motions ($V_{\mathrm{x}}^i$, $V_{\mathrm{y}}^i$) and their errors ($\sigma_{V\mathrm{x}}^i$, $\sigma_{V\mathrm{y}}^i$) are reported in table~\ref{maser_table}, together with their radial velocities $V_{\mathrm{LSR}}$, flux $S$, and detection S/Ns. 
In the case where a maser spot is detected in only two observations we do not define the errors for its motion and subsequently there is no error value written in the table. 

In figure~\ref{map}, we present the distribution and internal motions of the maser spots in the sky plane. 
They are distributed over an area of 110 mas $\times$ 50 mas corresponding to 26.0 AU $\times$ 16.3 AU at the source distance estimated in this work, i.e. 327\,pc. 
The distribution of maser spots is elongated along the East-West direction. 
At our estimated source distance a proper motion of 1 mas\,yr$^{-1}$ corresponds to a transverse velocity of 1.55 km\,s$^{-1}$. 
Scales for the magnitudes of the sky-plane velocity and size are presented at the top-right of the figure. 
The Color of each spot shows its $V_{\mathrm{LSR}}$, which ranges from $-$32 km\,s$^{-1}$(blue) to $-$23 km\,s$^{-1}$(red). 
The maser spot at $V_{\mathrm{LSR}} = -$29.73 km\,s$^{-1}$ (ID 57), which was used as a position reference in obtaining the internal motions, was placed at the map origin of figure~\ref{map}. 
Averaging over the relative internal motions of 42 maser spots, the typical transverse speed was obtained to be 2.71 mas\,yr$^{-1}$, corresponding to 4.20 km\,s$^{-1}$ at 327 pc. 

\begin{longtable}{ccccccccccc} 
 \caption{Parameters of the detected masers.}
 \label{maser_table}
 \hline 
 ID &$V_{\mathrm{LSR}}$& $X$& $Y$& $S$& S/N &$V_{\mathrm{x}}$& $\sigma_{V\mathrm{x}}$& $V_{\mathrm{y}}$& $\sigma_{V\mathrm{y}}$ & Grp. \\ 
$i$ &[km\,s$^{-1}$]& [mas]& [mas]& [Jy\,beam$^{-1}$]& &[mas\,yr$^{-1}$]&  & [mas\,yr$^{-1}$]&  & ID \\ 
 \hline \hline
 \endhead
 \hline
 \endfoot
 \hline
 \multicolumn{9}{@{}l@{}}{\hbox to 0pt{\parbox{155mm}{\footnotesize
\smallskip
Column~(1)---Component ID.
Column~(2)---LSR velocity in km\,s$^{-1}$. 
Column~(3)---offset positions in R.A. relative to the original phase center. 
Column~(4)---offset positions in Dec. relative to the original phase center.
Column~(5)---brightness of the spot at the first detection in Jy\,beam$^{-1}$. 
Column~(6)---signal-to-noise ratio (S/N). 
Column~(7)---best fit linear motion in R.A. in mas\,yr$^{-1}$.
Column~(8)---standard errors of the motions in R.A. 
Column~(9)---best fit linear motion in Dec in mas\,yr$^{-1}$.
Column~(10)---standard errors of the motions in Dec. 
Column~(11)---identification of maser group. 
 }}}
 \endlastfoot
1  & $-$23.20 & $-$23.74 & 27.46 & 3.1 & 17.3 & \dots & \dots & \dots & \dots & 6 \\ 
2  & $-$23.41 & $-$24.06 & 27.55 & 6.2 & 8.0 & 2.87 & 0.32 & 0.03 & 1.40 & 6 \\ 
3  & $-$23.41 & $-$23.11 & 26.08 & 2.6 & 13.7 & \dots & \dots & \dots & \dots & 6 \\ 
4  & $-$23.62 & $-$26.87 & 26.95 & 1.6 & 7.6 & \dots & \dots & \dots & \dots & 7 \\ 
5  & $-$23.62 & $-$25.85 & 30.69 & 1.8 & 8.8 & \dots & \dots & \dots & \dots & 7 \\ 
6  & $-$23.62 & $-$24.52 & 29.43 & 2.9 & 13.9 & 3.71 & \dots & $-$0.25 & \dots & 6 \\ 
7  & $-$23.62 & $-$23.90 & 27.74 & 18.2 & 87.1 & 3.05 & 0.90 & 0.66 & 0.09 & 6 \\ 
8  & $-$23.62 & $-$23.36 & 26.16 & 6.8 & 32.5 & \dots & \dots & \dots & \dots & 6 \\ 
9  & $-$23.83 & $-$26.35 & 28.94 & 4.5 & 8.5 & \dots & \dots & \dots & \dots & 7 \\ 
10 & $-$23.83 & $-$25.92 & 30.94 & 1.5 & 8.5 & \dots & \dots & \dots & \dots & 7 \\ 
11 & $-$23.83 & $-$24.52 & 29.65 & 2.5 & 14.3 & \dots & \dots & \dots & \dots & 6 \\ 
12 & $-$23.83 & $-$23.97 & 27.99 & 13.8 & 78.4 & 2.17 & 0.22 & 0.82 & 0.55 & 6 \\ 
13 & $-$23.83 & $-$23.20 & 26.25 & 2.2 & 12.6 & \dots & \dots & \dots & \dots & 6 \\ 
14 & $-$24.04 & $-$26.41 & 28.83 & 4.9 & 8.9 & 1.48 & \dots & $-$0.88 & \dots & 7 \\ 
15 & $-$24.04 & $-$24.05 & 28.23 & 8.0 & 52.8 & 2.46 & 0.24 & 0.40 & 0.74 & 6 \\ 
16 & $-$24.25 & $-$27.05 & 32.87 & 1.5 & 10.7 & \dots & \dots & \dots & \dots & 7 \\ 
17 & $-$24.25 & $-$26.45 & 29.30 & 4.7 & 8.0 & \dots & \dots & \dots & \dots & 7 \\ 
18 & $-$24.25 & $-$24.12 & 28.41 & 3.8 & 28.2 & 2.91 & 0.00 & $-$0.14 & 0.22 & 6 \\ 
19 & $-$24.46 & $-$27.13 & 32.98 & 8.0 & 50.1 & 1.96 & \dots & 1.50 & \dots & 7 \\ 
20 & $-$24.46 & $-$24.04 & 28.59 & 3.3 & 20.0 & 2.50 & \dots & 0.88 & \dots & 6 \\ 
21 & $-$24.67 & $-$27.78 & 34.89 & 2.3 & 10.7 & \dots & \dots & \dots & \dots & 7 \\ 
22 & $-$24.67 & $-$27.16 & 33.02 & 26.6 & 98.2 & 1.72 & 0.37 & 1.92 & 0.46 & 7 \\ 
23 & $-$24.67 & $-$26.31 & 31.14 & 2.5 & 9.3 & 0.06 & \dots & 3.09 & \dots & 7 \\ 
24 & $-$24.67 & $-$25.08 & 30.02 & 2.8 & 10.4 & 0.46 & 0.78 & 4.92 & 0.79 & 7 \\ 
25 & $-$24.88 & $-$27.74 & 34.88 & 4.3 & 13.8 & \dots & \dots & \dots & \dots & 7 \\ 
26 & $-$24.88 & $-$27.22 & 32.69 & 10.6 & 10.6 & 1.90 & 0.23 & 2.70 & 0.40 & 7 \\ 
27 & $-$24.88 & $-$26.31 & 31.39 & 3.7 & 11.9 & \dots & \dots & \dots & \dots & 7 \\ 
28 & $-$24.88 & $-$25.58 & 30.70 & 4.0 & 10.3 & 2.40 & 1.50 & 2.58 & 2.12 & 7 \\ 
29 & $-$25.10 & $-$27.98 & 34.82 & 3.5 & 12.4 & 4.98 & \dots & 0.23 & \dots & 7 \\ 
30 & $-$25.10 & $-$27.19 & 33.05 & 27.2 & 96.1 & 2.25 & 0.10 & 1.62 & 0.31 & 7 \\ 
31 & $-$25.10 & $-$25.81 & 30.96 & 3.0 & 10.7 & \dots & \dots & \dots & \dots & 7 \\ 
32 & $-$25.31 & $-$27.77 & 34.88 & 1.5 & 8.6 & \dots & \dots & \dots & \dots & 7 \\ 
33 & $-$25.31 & $-$27.19 & 33.05 & 8.4 & 55.9 & 1.73 & 0.20 & 2.13 & 0.32 & 7 \\ 
34 & $-$25.52 & $-$27.20 & 33.08 & 1.6 & 12.7 & 1.49 & \dots & 1.02 & \dots & 7 \\ 
35 & $-$26.99 & $-$16.03 & 26.61 & 21.5 & 28.5 & \dots & \dots & \dots & \dots & \dots \\ 
36 & $-$28.26 & 13.42 & 40.50 & 1.2 & 8.6 & \dots & \dots & \dots & \dots & 5 \\ 
37 & $-$28.47 & 13.40 & 40.42 & 2.0 & 14.4 & 2.91 & \dots & 1.34 & \dots & 5 \\ 
38 & $-$28.68 & 13.37 & 40.55 & 2.1 & 15.3 & 1.64 & \dots & $-$1.51 & \dots & 5 \\ 
39 & $-$28.89 & 0.01 & $-$0.02 & 1.3 & 9.8 & \dots & \dots & \dots & \dots & 4 \\ 
40 & $-$28.89 & 13.44 & 40.37 & 1.4 & 10.4 & \dots & \dots & \dots & \dots & 5 \\ 
41 & $-$29.10 & $-$1.69 & 0.74 & 1.6 & 9.3 & 2.36 & 0.78 & $-$2.59 & 1.00 & 4 \\ 
42 & $-$29.10 & $-$0.37 & $-$0.17 & 1.8 & 10.0 & 2.88 & 0.59 & $-$1.64 & 0.72 & 4 \\ 
43 & $-$29.31 & $-$94.25 & 11.49 & 1.2 & 8.0 & \dots & \dots & \dots & \dots & 1 \\ 
44 & $-$29.31 & $-$1.60 & 0.62 & 6.5 & 13.5 & 0.06 & \dots & $-$0.47 & \dots & 4 \\ 
45 & $-$29.31 & $-$0.26 & $-$0.02 & 6.7 & 23.6 & 1.45 & \dots & $-$1.10 & \dots & 4 \\ 
46 & $-$29.31 & $-$0.14 & 0.44 & 7.3 & 15.0 & \dots & \dots & \dots & \dots & 4 \\ 
47 & $-$29.31 & $-$0.02 & $-$0.31 & 12.5 & 25.8 & 1.01 & \dots & $-$1.15 & \dots & 4 \\ 
48 & $-$29.31 & 1.94 & $-$3.06 & 2.9 & 10.3 & \dots & \dots & \dots & \dots & 4 \\ 
49 & $-$29.52 & $-$94.22 & 12.36 & 2.4 & 13.2 & $-$2.64 & \dots & $-$1.52 & \dots & 1 \\ 
50 & $-$29.52 & $-$93.24 & 11.80 & 3.1 & 16.8 & $-$1.78 & 0.84 & $-$2.93 & 1.46 & 1 \\ 
51 & $-$29.52 & $-$67.58 & $-$0.40 & 3.3 & 8.2 & $-$3.23 & 0.14 & $-$1.33 & 1.41 & 2 \\ 
52 & $-$29.52 & $-$0.18 & $-$0.37 & 3.2 & 14.1 & \dots & \dots & \dots & \dots & 4 \\ 
53 & $-$29.52 & $-$0.04 & $-$0.04 & 20.6 & 50.7 & \dots & \dots & \dots & \dots & 4 \\ 
54 & $-$29.52 & 0.06 & $-$0.12 & 2.4 & 13.2 & \dots & \dots & \dots & \dots & 4 \\ 
55 & $-$29.73 & $-$93.41 & 12.02 & 7.5 & 31.6 & $-$3.97 & 0.16 & $-$0.30 & 0.17 & 1 \\ 
56 & $-$29.73 & $-$67.59 & $-$0.49 & 7.6 & 23.3 & $-$3.31 & 0.26 & $-$0.62 & 0.96 & 2 \\ 
57 & $-$29.73 & 0.00 & 0.00 & 2.6 & 16.4 & 1.01 & \dots & $-$0.88 & \dots & 4 \\ 
58 & $-$29.73 & 1.17 & 3.57 & 5.1 & 11.1 & \dots & \dots & \dots & \dots & 4 \\ 
59 & $-$29.73 & 2.09 & $-$2.42 & 2.1 & 7.7 & \dots & \dots & \dots & \dots & 4 \\ 
60 & $-$29.94 & $-$93.73 & 12.19 & 6.2 & 20.1 & $-$4.58 & 1.91 & $-$1.22 & 1.79 & 1 \\ 
61 & $-$29.94 & $-$67.57 & $-$0.50 & 7.8 & 27.4 & $-$3.15 & 0.07 & $-$0.27 & 0.65 & 2 \\ 
62 & $-$29.94 & 0.00 & 0.00 & 4.2 & 7.9 & \dots & \dots & \dots & \dots & 4 \\ 
63 & $-$29.94 & 1.45 & $-$7.30 & 3.9 & 10.1 & \dots & \dots & \dots & \dots & 4 \\ 
64 & $-$29.94 & 2.25 & $-$0.81 & 3.0 & 10.6 & 0.19 & 0.34 & 0.00 & 0.04 & 4 \\ 
65 & $-$29.94 & 4.24 & $-$3.94 & 3.8 & 13.4 & \dots & \dots & \dots & \dots & 4 \\ 
66 & $-$30.15 & $-$94.20 & 12.29 & 3.1 & 10.7 & $-$3.28 & \dots & $-$0.42 & \dots & 1 \\ 
67 & $-$30.15 & $-$67.56 & $-$0.55 & 3.6 & 11.2 & \dots & \dots & \dots & \dots & 2 \\ 
68 & $-$30.15 & 0.06 & $-$0.25 & 8.2 & 16.3 & \dots & \dots & \dots & \dots & 4 \\ 
69 & $-$30.15 & 2.27 & $-$1.14 & 6.6 & 14.9 & 0.87 & \dots & $-$0.74 & \dots & 4 \\ 
70 & $-$30.15 & 2.29 & $-$0.41 & 8.9 & 20.3 & 1.28 & \dots & $-$1.02 & \dots & 4 \\ 
71 & $-$30.15 & 2.41 & $-$0.79 & 7.2 & 22.5 & 1.89 & 0.13 & $-$1.24 & 0.06 & 4 \\ 
72 & $-$30.36 & $-$94.14 & 12.33 & 1.5 & 8.1 & $-$4.10 & \dots & $-$0.06 & \dots & 1 \\ 
73 & $-$30.36 & 0.00 & 0.00 & 8.7 & 27.8 & \dots & \dots & \dots & \dots & 4 \\ 
74 & $-$30.36 & 2.55 & $-$0.84 & 5.9 & 22.0 & $-$0.69 & 0.93 & $-$0.64 & 1.46 &4  \\ 
75 & $-$30.57 & 0.40 & $-$0.42 & 2.9 & 7.0 & \dots & \dots & \dots & \dots & 4 \\ 
76 & $-$30.57 & 2.32 & $-$0.62 & 3.2 & 13.0 & \dots & \dots & \dots & \dots & 4 \\ 
77 & $-$30.57 & 2.98 & $-$1.20 & 2.9 & 11.7 & \dots & \dots & \dots & \dots & 4 \\ 
78 & $-$30.57 & 3.50 & $-$1.58 & 3.3 & 10.4 & \dots & \dots & \dots & \dots & 4 \\ 
79 & $-$30.99 & $-$0.03 & $-$0.08 & 1.1 & 7.9 & \dots & \dots & \dots & \dots & 4 \\ 
80 & $-$31.21 & $-$66.53 & 19.62 & 1.7 & 12.8 & \dots & \dots & \dots & \dots & 3 \\ 
81 & $-$31.42 & $-$66.30 & 19.19 & 1.3 & 9.9 & $-$1.45 & \dots & 0.85 & \dots & 3 \\ 
82 & $-$31.63 & $-$66.38 & 19.14 & 2.4 & 16.9 & $-$1.72 & \dots & 1.04 & \dots & 3 \\ 
83 & $-$31.84 & $-$66.47 & 19.13 & 2.1 & 15.0 & $-$1.27 & \dots & 0.31 & \dots & 3 \\ 
84 & $-$32.05 & $-$66.76 & 19.20 & 2.2 & 17.2 & \dots & \dots & \dots & \dots & 3 \\ 
\end{longtable}

\begin{table*}[!tb]
\begin{center}
\caption{Maser Groups.}
\label{grp_table}
\begin{tabular}{crrrrrr} 
\hline
 Grp.ID & $V_{\mathrm{LSR}}^{\mathrm{Grp}}$  & $\Delta V_{\mathrm{LSR}}$\footnotemark[$\dag$] & $X^{\mathrm{Grp}}$   & $Y^{\mathrm{Grp}}$   & $V_{\mathrm{x}}^{\mathrm{Grp}}$ & $V_{\mathrm{y}}^{\mathrm{Grp}}$ \\ 
  $j$  & [km\,s$^{-1}$]     & [km\,s$^{-1}$]           & [mas] & [mas] & [mas\,yr$^{-1}$] & [mas\,yr$^{-1}$] \\ 
 \hline \hline 
1 & $-$29.79 & 1.05 & $-$93.88 &   12.07 & $-$3.39 & $-$1.08 \\
2 & $-$29.84 & 0.63 & $-$67.58 & $-$0.49 & $-$3.23 & $-$0.74 \\
3 & $-$31.63 & 0.84 & $-$66.49 &   19.26 & $-$1.48 &    0.73 \\
4 & $-$29.85 & 2.10 &     0.95 & $-$0.71 &    1.13 & $-$1.06 \\
5 & $-$28.58 & 0.63 &    13.41 &   40.46 &    2.28 & $-$0.09 \\
6 & $-$23.76 & 1.26 & $-$23.88 &   27.80 &    2.81 &    0.34 \\
7 & $-$24.62 & 2.90 & $-$26.74 &   31.82 &    1.86 &    1.89 \\
\hline 
  \multicolumn{7}{@{}l@{}}{\hbox to 0pt{\parbox{110mm}{\footnotesize
  \smallskip
 \par\noindent
 \footnotemark[$\dag$] $\Delta V_{\mathrm{LSR}}$ indicates a difference between maximum and minimum $V_{\mathrm{LSR}}$ values within each maser group. 
     }\hss}}
\end{tabular} 
\end{center}  
\end{table*}

\section{Discussion} 
\label{discussion}
\subsection{Systemic motion of T~Lep}
\label{star_position}
From our VLBI observations, we cannot detect any radiation from the central stellar photosphere, and thus we cannot obtain the exact position of the central star directly. 
However, using the distribution and kinematics of the maser spots, we can estimate it. 
As shown in the previous section, we consider that the distributions and kinematics of masers around T~Lep are better traced by consideration of maser groups, rather than considering all maser spots individually.
In this section we therefore denote the ``group position'' ($X^{\mathrm{Grp}}, Y^{\mathrm{Grp}}$) and ``group velocity'' ($V_{\mathrm{x}}^{\mathrm{Grp}}, V_{\mathrm{y}}^{\mathrm{Grp}}$) as positions and velocities averaged over all maser spots in each group, these are also presented in table~\ref{grp_table}. 
We also show the ``group radial velocity" $V_\mathrm{LSR}^\mathrm{Grp}$ which is the radial velocity averaged over all maser spots in each group, and radial velocity difference in each group $\Delta V_\mathrm{LSR}$, in table~\ref{grp_table}. 

There are a few possible methods to estimate the position of the central star from group positions.  
At first, we simply took the median of the maser group positions in R.A. and Dec. 
This lead to an estimated position of ($-40.24\pm36.52$, $19.88\pm14.78$) mas and is shown by a cross drawn with a solid line (line 1) in figure~\ref{map_grp}, where error bars come from the standard deviations of all group positions with respect to the median. 
Secondly, we took the averages of the group positions. 
This approach lead to an estimated position of ($-37.74\pm36.44$, $18.60\pm14.72$) mas and is shown as a cross drawn with a dashed line (line2) in figure~\ref{map_grp}, where error bars are the standard deviations of all group positions with respect to the average. 
In addition to the group positions, we considered the internal motion of each maser group. 
If all maser groups move radially outwards from the central star, motion vectors of maser groups should intersect at a common origin. 
Using the observed motion vectors of maser groups on the sky plane, we noted all intersection points involving any two of the seven maser groups. 
Out of 21 intersection points we excluded 9 points which were located beyond the maser distribution. 
The average of these intersection points was ($-41.80\pm17.57$, $20.34\pm9.94$) mas and is shown in figure~\ref{map_grp} as a two-dot chain line (line 3).

These three expected positions of the central star were consistent within their errors. 
We therefore concluded the position of the central star to be ($X^{\ast}$, $Y^{\ast}$) $=$ ($-39.93\pm30.18$, $19.61\pm13.15$) mas which is shown in figure~\ref{map} as a white cross.
The final positional error was estimated as the geometric average of the errors from the above three procedures.  
Subtracting the mean motion 
($V_{\mathrm{x}}^{\mathrm{mean}}$, $V_{\mathrm{y}}^{\mathrm{mean}}$) 
from the linear proper motion of the maser spot used in parallax estimation (ID 57), we also obtained a systemic velocity of T~Lep ($V_\mathrm{x}^{\ast}$, $V_\mathrm{y}^{\ast}$), which we estimate to be  (14.60$\pm$0.50, $-$35.43$\pm$0.79) mas\,yr$^{-1}$. 
We use these motions in section~\ref{3D-kinematic} in order to discuss kinematics of T~Lep in a Galacto-centric coordinate frame. 

In the first Hipparcos catalog~\citep{per97} and revised one~\citep{van07}, the proper motion of T~Lep was reported to be ($8.17\pm1.63$, $-31.63\pm1.74$) mas\,yr$^{-1}$ and 
($12.31\pm1.18$, $-33.31\pm1.10$) mas\,yr$^{-1}$, respectively. 
The proper motions and parallaxes from the aforementioned works, along with our results, are shown in table~\ref{param}.
The parallax from our measurement improved the accuracy of the distance to T~Lep, and the consistency of proper motion estimations between \citet{per97}, \citet{van07}, and  our measurement supports the validity of our method in investigating the internal motions of maser spots, and the position of the central star. 

\begin{table*}[!tb]
\begin{center}
\caption{Astrometric Parameters of T Lep.}
\label{param}
\begin{tabular}{cccccc} 
\hline 
& $\mu_{\mathrm{X}}$ & $\mu_{\mathrm{Y}}$  & Parallax & Reference & \\ 
& [mas\,yr$^{-1}$]  & [mas\,yr$^{-1}$]     & [mas]           &  & \\ 
 \hline \hline 
&14.60$\pm$0.50 & $-$35.43$\pm$0.79 & 3.06$\pm$0.04 & This paper. & \\
&8.17$\pm$1.63 & $-$31.63$\pm$1.74 &  $-$1.77$\pm$2.73 & \cite{per97} (HIPPARCOS) & \\ 
&12.31$\pm$1.18 & $-$33.31$\pm$1.10 & 1.37$\pm$1.44 & \cite{van07} (Revised HIPPARCOS) & \\
\hline 
\end{tabular}
\end{center}
\end{table*}

\subsection{Superposition of radio and infrared images}
\label{map_superpose}
Today, the consensus regarding the global picture of AGB stars is an onion-like structure with molecular shells and dust shells surrounding the photosphere (e.g. \cite{wit07}). 
This kind of structure is confirmed in radio observations. 
For example, \citet{rei97} observed Mira and semiregular variables with the VLA. 
They resolved the stellar disk of W~Hya and suggested that there exists a radio photosphere encompassing the stellar disk. 

Recently, infrared interferometers are playing an important role in the investigation of the circumstellar environment of AGB stars (e.g. \cite{leb09, wit05, wit07}). 
Using the Very Large Telescope Interferometer (VLTI, \cite{hag08}), they created infrared images with resolutions as fine as those of radio interferometers. 
\citet{leb09} observed T~Lep with the VLTI at $J$-, $H$-, and $K$-bands and obtained images and sizes of the central star with its surrounding shell. 
This represented the first direct evidence of the spherical morphology of the photosphere and molecular shell in T~Lep. 
The images were taken in a short period, like a snapshot, and the sizes are thus considered to be applicable explicitly to that time.
Although they revealed the stellar surroundings, the linear size had remained uncertain since conversion from angular size to linear size requires accurate knowledge of the source distance. 
We estimate the linear sizes of the photosphere and surroundings based on our distance measurement. 
Prior to this work \citet{leb09} estimated the linear size of T~Lep using a parallax of $5.97\pm0.70$ mas. 
However, this was in fact the value for another star and the actual parallax of $1.37\pm1.44$ mas from the revised Hipparcos catalog \citep{van07} has too large an error to estimate the distance reliably. 

In figure~\ref{map}, we overlaid the VLTI image of T~Lep from \citet{leb09} with H$_2$O masers from this work using the estimated stellar position obtained in section~\ref{star_position}. 
Using our distance of 327 pc, the radius of the molecular shell of 7.5 mas \citep{leb09} corresponds to 2.45 AU ($=$ 527 $R_{\odot}$), and photospheric radius of 2.9 mas \citep{leb09} corresponds to 0.95 AU ($=$ 204 $R_{\odot}$).  
The projected distance of maser spots from the central star was seen to be 36 mas, or 12 AU, which is 12 times larger than the radius of the stellar photosphere. 
Using the same instrument as \citet{leb09}, \citet{wit07} observed the Mira variable S~Ori and reported that the stellar photosphere of S~Ori shows a phase-dependent variability with an amplitude of $\sim$20\,\% that is well related in phase with the visual lightcurve. 
\citet{zha11} also reported a variability of the photosphere of the semiregular variable W~Hya. 
The photosphere of T~Lep has not been continuously observed like S~Ori or W~Hya, monitoring observations with infrared interferometer and accurate distances would be helpful for a better understanding of linear scale properties of Mira variables. 

\subsection{Calibration of the period-luminosity relation}
\label{section_plr}
By analyzing  AGB stars in the Large Magellanic Cloud (LMC), \citet{woo00} found at least five distinct period-luminosity relation (PLR) sequences. 
\citet{ita04} also confirmed these relations using larger number of stars. 
Thanks to the large number of samples, the PLR in LMC is well established. 
However, the PLR for Mira variables in our Galaxy is yet to be well established because only a few Mira variables have accurately measured distances, using reliable methods. 
In determining the Galactic PLR, we have to convert an apparent magnitude $m_K$ to an absolute magnitude $M_K$ based on the source distance, therefore, the accuracy of the distance estimate is crucial for this study. 
The sample number is also of influence, since large samples contribute to make the Galactic PLR more accurate. 

The PLR has been investigated previously by \citet{van97} using 16 sources from first Hipparcos catalog \citep{per97}, however uncertainties in $M_K$ were very large due to large distance errors. 
More recently, \citet{whi08} compiled parallaxes from VLBI observations together with the revised Hipparcos Catalog \citep{van07} and reported the Galactic PLR.  
Recent VLBI observations with VERA give the parallaxes of four variables; 
2.33$\pm$0.13 mas for S~Crt \citep{nak08}, 
4.7$\pm$0.8 mas for R~Aqr \citep{kam10}, 
0.75$\pm$0.03 mas for SY~Scl \citep{nyu11}, and 
7.31$\pm$0.50 mas for RX~Boo \citep{kam12}. 
In this paper, we provide the addition of a new source T~Lep, with its parallax of 3.06$\pm$0.04 mas, to the sample set (table~\ref{parallax_vera3}). 
Using the distance $D=$327$\pm 4.3$ pc of T~Lep, the apparent $K$ magnitude $m_{K}=0.12$ (Fourier mean magnitude in \cite{whi00}) can be converted to an absolute magnitude $M_K=-7.45 \pm 0.03$, where the error of 0.03 mag is attributed to the parallax error. 
In table~\ref{parallax_vera3} we also give $M_K$ for the other sources S~Crt, R~Aqr, SY~Scl, and RX~Boo.  
To derive $M_K$ for these four sources, we used values of $m_{K}$ from previously published papers, and distances determined with VERA. 
The variability types (Mira or semiregular; SR), parallaxes, pulsation periods,  $m_{K}$, and  $M_{K}$ are summarized in table~\ref{parallax_vera3}, and references of the parallaxes and $m_{K}$ are also presented in the footnote of the table. 
In addition to the results from VERA, we compiled data from five published sources (S~CrB, U~Her, RR~Aql, W~Hya, and R~Cas) whose parallaxes were measured with astrometric VLBI observations using the Very Long Baseline Array (VLBA), and we presented them also in the lower half of table~\ref{parallax_vera3}. 
We estimated the $M_{K}$ errors of all sources assuming that the errors can be attributed to only the distance uncertainties.

\begin{figure}[tb]
\begin{center}
 \includegraphics[width=87mm, angle=0]{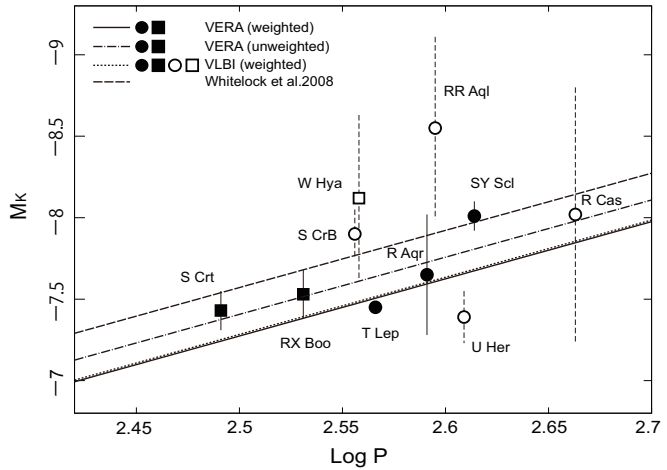}
\end{center}
 \caption{ 
Period$-$luminosity relation derived from astrometric VLBI observations. 
Filled symbols represent absolute magnitudes $M_K$ that derived from VERA observations. 
Open symbols represent those from other VLBI observations conducted by \citet{vle03} and \citet{vle07}. 
Square symbols are used to denote semiregular variables. 
Dashed line shows a relation reported by \citet{whi08}. 
}
\label{plr}
\end{figure}

We define the PLR in the form of 
$ M_K = -3.51 \, \mathrm{log}P +  \delta$. 
Since it is difficult to determine the slope of the PLR using a small number of sample, we adopted the slope of $-3.51$ which was obtained from the relation in LMC \citep{whi08}. 
We solve $\delta$ through a weighted least-squares fitting. 
Using all ten sources on table~\ref{parallax_vera3}, $\delta$ was obtained to be 1.49$\pm$0.05, and this relation is presented with the dotted line in figure~\ref{plr}. 
When we use five sources (T~Lep, S~Crt, R~Aqr, SY~Scl, RX~Boo) whose distances were determined with VERA, $\delta$ was obtained to be 1.50$\pm$0.06, and this relation is presented with the solid line in figure~\ref{plr}.  
For a comparison between our study and previous works, we also show the relation reported by \citet{whi08} in figure~\ref{plr} using a dashed line. 
Fitting results from two sample sets  (all ten sources and five VERA sources) become close to each other because the $M_{K}$ error of T~Lep is very small and a weighting of this source is large in the fittings. 
Using the five sources from VERA observations, we also solved $\delta$ through an unweighted least-squares fitting in order to avoid the effect of large weighting of T~Lep. 
The $\delta$ was obtained to be 1.37$\pm$0.07, and this relation is presented with one-dotted chain line in figure~\ref{plr}. 
Every $\delta$ values obtained from the fittings are larger compared to the relation reported by \citet{whi08}. 
At present, accuracies of the distances largely differ for each source, and the differences in the estimation of $\delta$ should be carefully addressed. 
For a better understanding of the relation, we are monitoring more Mira variables with VERA to increase sample numbers and to obtain more accurate values of $M_K$. 

\begin{table*}[!tb]
\caption{Results from VLBI astrometry.}
\label{parallax_vera3}
\begin{center}
\begin{tabular}{ccccccc} 
\hline
Source & Type & Parallax\footnotemark[$\dag$] & $P$ & $\mathrm{Log}P$ & $m_K$\footnotemark[$\ddag$] & $M_K$ \\
          &  & [mas] & [day] &                          & [mag]  & [mag]  \\ \hline \hline
T~Lep & Mira & 3.06$\pm$0.04$^{\mathrm{(a)}}$ & 368 & 2.566 & 0.12$^{\mathrm{(h)}}$ & $-7.45\pm0.03$ \\ 
S~Crt & SR & 2.33$\pm$0.13$^{\mathrm{(b)}}$ & 310\footnotemark[$\ast$] & 2.190 & 0.73$^{\mathrm{(i)}}$ & $-7.43\pm0.12$ \\ 
R~Aqr & Mira & 4.7$\pm$0.8$^{\mathrm{(c)}}$ & 390 & 2.591 & $-$1.01$^{\mathrm{(h)}}$ & $-7.65\pm0.37$ \\
SY~Scl & Mira & 0.75$\pm$0.03$^{\mathrm{(d)}}$ & 411 & 2.614 & 2.61$^{\mathrm{(j)}}$ & $-8.01\pm0.09$ \\ 
RX~Boo & SR & 7.31$\pm$0.5$^{\mathrm{(e)}}$ & 340 & 2.531 & $-$1.85$^{\mathrm{(k)}}$ & $-7.53\pm0.15$ \\ \hline
S~CrB & Mira & 2.39$\pm$0.17$^{\mathrm{(f)}}$ & 360 & 2.556 & 0.21$^{\mathrm{(h)}}$ & $-7.90\pm0.15$ \\
U~Her & Mira & 3.76$\pm$0.27$^{\mathrm{(f)}}$ & 406 & 2.609 & $-$0.27$^{\mathrm{(h)}}$ & $-7.39\pm0.16$ \\
RR~Aql & Mira & 1.58$\pm$0.40$^{\mathrm{(f)}}$ & 394 & 2.595 & 0.46$^{\mathrm{(h)}}$ & $-8.55\pm0.56$ \\
W~Hya & SR & 10.18$\pm$2.36$^{\mathrm{(g)}}$ & 361 & 2.558 & $-$3.16$^{\mathrm{(h)}}$ & $-8.12\pm0.51$ \\
R~Cas & Mira & 5.67$\pm$1.95$^{\mathrm{(g)}}$ & 460 & 2.663 & $-$1.79$^{\mathrm{(l)}}$ & $-8.02\pm0.78$ \\
\hline 
  \multicolumn{7}{@{}l@{}}{\hbox to 0pt{\parbox{115mm}{\footnotesize
  \smallskip
 \par\noindent
 \footnotemark[$\dag$]
Reference of the parallax; 
$^{\mathrm{(a)}}$This paper, 
$^{\mathrm{(b)}}$\cite{nak08}, 
$^{\mathrm{(c)}}$\cite{kam10}, 
$^{\mathrm{(d)}}$\cite{nyu11}, 
$^{\mathrm{(e)}}$\cite{kam12}, 
$^{\mathrm{(f)}}$\cite{vle07}, and  
$^{\mathrm{(g)}}$\cite{vle03}. \\
 \footnotemark[$\ddag$]Reference of the $m_K$; 
$^{\mathrm{(h)}}$\cite{whi00} (Fourier mean magnitude), 
$^{\mathrm{(i)}}$\cite{jul92}, 
$^{\mathrm{(j)}}$\cite{whi94}, 
$^{\mathrm{(k)}}$\cite{gla07}, and
$^{\mathrm{(l)}}$\cite{fea00}.  \\
 \footnotemark[$\ast$] For the period of S~Crt, we use 310~days, which is the double of its first overtone period of 155~day. 
     }\hss}}
\end{tabular} 
\end{center}  
\end{table*}   

\subsection{Galactocentric kinematics}
\label{3D-kinematic}
From our astrometric observations, the three-dimensional positions and velocities of T~Lep were revealed. 
Using standard assumptions about the galactic rotation and solar motion, we can convert the observed motion in the helio-centric coordinate to motions in the Galacto-centric coordinate. 
In this section we derive the non-circular motion (peculiar motion) of T~Lep.
Here, velocity components of the peculiar motion of T~Lep (U$^\ast$, V$^\ast$, W$^\ast$) are defined at the source local position.
U$^\ast$ is the velocity radially toward to the Galactic center, 
V$^\ast$ is in the direction of the Galactic circular rotation, 
and W$^\ast$ is in the direction to the North Galactic Pole, perpendicular to the Galactic plane. 
In the conversion procedure, we adopted $\Theta_0 =$ 220 km\,s$^{-1}$ (IAU recommendation) as the circular rotation velocity of the LSR. 
The same value was used as T~Lep's circular rotation velocity. 
We also adopted a value of 8.5 kpc (IAU recommendation) for the Galacto-centric distance of the Sun.
In a detailed study by \citet{fra09},  the solar motion was reported to be  (U$_\odot$, V$_\odot$, W$_\odot$) $=$ (7.5$\pm$1.0, 13.5$\pm$0.3, 6.8$\pm$0.1)\,km\,s$^{-1}$, we used these values in our analysis. 

In the calculation of the three-dimensional kinematics of T~Lep, we used the systemic proper motion of ($V_\mathrm{x}^{\ast}$, $V_\mathrm{y}^{\ast}$)$=$(14.60$\pm$0.50, $-$35.43$\pm$0.79) mas\,yr$^{-1}$ 
and the LSR velocity $-$27.63 km\,s$^{-1}$ which was the center velocity of the maser spots shown in table~\ref{maser_table}.
The resultant peculiar motion of T~Lep was obtained to be (U$^\ast$, V$^\ast$, W$^\ast$) $=$ (57.96, $-$36.14, 14.67)\,km\,s$^{-1}$. 
This shows a large discrepancy from the assumed circular rotation, indicating an inward motion to the Galactic center, a large delay from the circular rotation, and a northward motion from the Galactic plane. 
However, the individual peculiar motion of T~Lep is consistent with the results published in \citet{fea00} which also indicates inward motions to the Galactic center and a delay from the Galactic rotation, which represents a different trend from the statistic properties of variable stars with similar periods in \citet{fea00}. 
The obtained peculiar motion of T~Lep is indicated with a white arrow at the top-right of figure~\ref{map}. 
The direction parallel to the Galactic plane is also shown with another white arrow at same position in figure~\ref{map}, where the Galactic longitude $l$ increases toward the south-east of this map. 

If we assume that the inter stellar medium (ISM) is fixed with regards to local circular rotation, the peculiar motion can be treated as a source motion relative to the ISM. 
In this picture, the linear velocity of T~Lep relative to the ISM is obtained to be 69.86\,km\,s$^{-1}$, which is the root sum square of the three components of peculiar motion. 
Recently, evidence of the interaction between the ISM and circumstellar medium (CSM) of evolved stars was detected in infrared images \citep{cox12,dec12}. 
Clear images of bow shocks in many sources are reported and these bow shocks often appear in front of the source proper motion. 
Since T~Lep is an evolved star with a high mass-loss rate, we can expect a bow shock or dust shell around T~Lep. 
Up to now, there is no publication similar to the works of \citet{cox12} or \citet{dec12} showing the infrared image around T~Lep. 
In the near future, we expect the detection of a bow shock that indicates clear consistency with our kinematic estimation. 

\section{Summary} 
\label{summary}
Phase referencing VLBI observations with VERA towards T~Lep allowed for the evaluation of a precise distance estimate, based on the measurement of annual parallax. 
The obtained parallax of 3.06$\pm$0.04\,mas corresponds to a distance of 327$\pm$4.3 pc. 
We revealed the distribution of maser spots in an area of 110 mas $\times$ 50 mas, and the internal kinematics of maser spots showed an outward motion from the distribution center.
Using the obtained systemic proper motion of T~Lep  ($V_\mathrm{x}^{\ast}$, $V_\mathrm{y}^{\ast}$) $=$ (14.60$\pm$0.50, $-$35.43$\pm$0.79) mas\,yr$^{-1}$ and LSR velocity of $-$27.63 km\,s$^{-1}$, we deduced a peculiar motion of (U$^\ast$, V$^\ast$, W$^\ast$) $=$ (57.96, $-$36.14, 14.67)\,km\,s$^{-1}$, assuming circular rotation at the source local position.  
This shows a large discrepancy from the assumed circular rotation. 

We superimposed an image of the H$_2$O masers from this work with an interferometric infrared image of T~Lep.
For the position of the central star, we assumed an estimated position from our maser distribution and kinematic analysis. 
Using the obtained distance of 327 pc, the angular radius of the molecular shell (7.5 mas) and photosphere (2.9 mas) revealed with the VLTI by \citet{leb09} were converted to the linear scales of 2.45 AU and 0.95 AU, respectively.   

For the determination of the Galactic PLR, we compiled published results obtained with astrometric VLBI observations. 
Using ten sources including our contribution of ``T~Lep", we calibrated the PLR for Galactic Mira variables in the form $M_K = -3.51 \, \mathrm{log}P + \delta$ by solving the parameter $\delta$ 
via weighted least-squares fitting. 
The $\delta$ was obtained to be 1.49$\pm$0.05. 
We are continuing astrometric VLBI observation towards Galactic Mira variables to calibrate the PLR more accurately. 

\vspace{5mm}
First author would like to thank R.~Burns for careful reading of the manuscript, and helpful advices. 
We gratefully acknowledge T.~Jike and T.~Kurayama who developed software packages which aided us in reduction of the astrometric data.

\end{document}